# Evidence of topological Shiba bands in artificial spin chains on superconductors

**Lucas Schneider[1], Philip Beck[1], Thore Posske[2,3], Daniel Crawford[4], Eric Mascot[5], Stephan Rachel[4], Roland Wiesendanger[1] and Jens Wiebe[1,*]**

[1]Department of Physics, Universität Hamburg, D-20355 Hamburg, Germany.

[2]I. Institute for Theoretical Physics, Universität Hamburg, D-20355 Hamburg, Germany.

[3]Centre for Ultrafast Imaging, Luruper Chaussee 149, D-22761 Hamburg, Germany.

[4]School of Physics, University of Melbourne, Parkville, VIC 3010, Australia.

[5]Department of Physics, University of Illinois at Chicago, Chicago, IL 60607, USA

[*]E-mail: jwiebe@physnet.uni-hamburg.de

## Abstract
A major challenge in developing topological superconductors for implementing topological quantum computing is their characterization and control. It has been proposed that a *p*-wave gapped topological superconductor can be fabricated with single-atom precision by assembling chains of magnetic atoms on *s*-wave superconductors with spin-orbit coupling. Here, we analyze the Bogoliubov quasiparticle interference in atom-by-atom constructed Mn chains on Nb(110) and for the first time reveal the formation of multi-orbital Shiba bands using momentum resolved measurements. We find evidence that one band features a topologically non-trivial *p*-wave gap as inferred from its shape and particle-hole asymmetric intensity. Our work is an important step towards a distinct experimental determination of topological phases in multi-orbital systems by bulk electron band structure properties only.

## Main
Topological superconductors (TSC) in one dimension can host zero-energy Majorana bound states (MBS) at their edges[1–6], which are protected from excitations by a topological, e.g. *p*-wave, gap. Due to their non-Abelian exchange statistics, they are candidates for topological qubits, offering ways to strong stability against decoherence[7–9]. With the hope to realize TSC, one-dimensional magnetic nanostructures proximity-coupled to *s*-wave superconductors[10–16] have been intensively studied by scanning tunneling spectroscopy (STS). The various experimentally investigated systems include ferromagnetic Fe[17–21] and Co[22] nanowires on Pb(110) and chains of Fe atoms with helical spin order on Re(0001) that can be tailored and controlled down to each individual atom[23,24]. In case of ferromagnetically ordered nanowires, topologically non-trivial phases are predicted to be realized in the presence of Rashba spin-orbit coupling whenever the number of spin-polarized bands crossing the Fermi energy $E_\mathrm{F}$ - in the absence of superconducting pairing - is odd[11]. The Rashba spin-orbit coupling (SOC) together with the superconducting pairing then induces gaps $\Delta_\mathrm{ind}$ of *p*-wave nature in these bands. For a weak interaction between the orbitals of the magnetic atoms in the wire[10], the low energy electronic Shiba bands are dominated by the hybridizing Yu-Shiba-Rusinov (YSR) states of the individual atoms[25–31]. Experimental STS studies of these systems typically argued with the presence or absence of zero energy edge modes[17–24]. This way of identifying a TSC phase bears the risk of confusing trivial edge states, which are ubiquitous in such systems, with MBSs[20,24]. Moreover, such studies concentrated on the consequence of, rather than the cause for TSC, which is purely determined by the bulk band structure in the interior of the chain. In order to gain a better understanding of magnet-superconductor hybrid systems, experimental measurements of the bands responsible for the formation of MBS would be highly desirable. Unfortunately, it is extremely difficult to perform standard momentum ($k$)-resolved measurements, like angle-resolved photoemission spectroscopy, on nano-scale or even atomic chains on surfaces because of their sparse distribution, length variations, or the impossibility to relocate chains which have been built atom-by-atom using scanning tunneling microscope (STM) tip-induced atom manipulation[23,24]. However, in principle, it is possible to extract the band structure from quasiparticle interference (QPI)[32–37] of the electronic states inside the chain. Scattering of quasiparticles with energy $E$ at defects or at the chain's edges between initial and final momenta $\mathbf{k}_i$ and $\mathbf{k}_f$, respectively, leads to interference and thereby produces modulations of the local electron density of states (LDOS) with wavelength $\tilde{\lambda}$ which can be directly imaged using STS. The resulting dispersions of the scattering vectors $|\mathbf{q}| = |\mathbf{k}_f - \mathbf{k}_i| = \frac{2\pi}{\tilde{\lambda}}$ are closely related to the quasiparticle band structure. While this technique has been applied successfully to a variety of complex electron systems as in high-temperature superconducting cuprates[36], Fe-based superconductors[38], heavy Fermions[39], or topological insulators[40], it has not been used so far for experimentally revealing the band structure of magnetic chains coupled to superconductors.

Here, we perform QPI imaging on ferromagnetic manganese (Mn) chains on the elemental superconductor niobium (Nb), a candidate for realizing one-dimensional TSC. The Mn chains have been assembled by STM-based atom manipulation on the clean and non-reconstructed (110) surface of Nb. By changing the length in an atom-by-atom fashion, from a single Mn atom to chains comprising up to several tens of Mn atoms, we were able to observe the multi-orbital Shiba band formation inside the energy gap $\varDelta_s$ of the superconducting Nb substrate, starting from the single impurity YSR states[26], and experimentally study the Shiba bands in view of their topological properties.

**Multi-orbital Yu-Shiba-Rusinov states of Mn atoms on superconducting Nb**
We use a clean Nb(110) crystal as a substrate for the chains[41] (see Methods & Supplementary Note 1), exhibiting the largest gap $\varDelta_s$ = 1.51 meV[42] among all elemental superconductors and therefore providing an improved effective energy resolution[43] with respect to previous experiments[23,24]. To enhance our energy resolution further[29], we indented our W tip into the substrate for several nanometers such that a superconducting cluster is formed on the tip apex. Subsequently, we deposited single Mn atoms, appearing as ≈ 90 pm high protrusions in Fig. 1a. We measure clear superconductor-insulator-superconductor (SIS) tunneling (gray spectrum in the upper panel of Fig. 1d) on the bare Nb substrate with the Nb covered tip: All features of the sample's LDOS appear shifted by an energy given by the gap of the superconducting tip, $\varDelta_t$ = 1.42 meV (Supplementary Note 3). Therefore, the coherence peak of Nb is visible at $\pm(\varDelta_s+\varDelta_t)$. Measuring tunneling spectra on top of the Mn atoms (red spectrum) reveals four pairs of additional peaks inside this gap, named $\alpha_\pm, \beta_\pm, \gamma_\pm, \delta_\pm$, where the peaks above (+) and below (-) $E_F$ appear at bias voltages symmetrically around zero. They are assigned to the multiplet of YSR states of the Mn atom due to the five $d$ orbitals that induce the magnetism[30,31]. We numerically deconvolute the measured SIS spectra[44] in the following (bottom panel of Fig. 1d) such that the presented data directly corresponds to the LDOS of the sample as it is common for STS experiments (Supplementary Note 2). The spatial structure of the YSR states can be determined using d$I$/d$V$ maps (see Methods) at the peak energies (Fig. 1e) which resemble the associated shapes of the according orbitals to good approximation[30,31]. Thereby, we attribute the state $\alpha$, which is strongest in intensity and closest to the coherence peak, to the atomic $d_{z^2}$ orbital, the $\beta$-state to the in-plane $d_{xy}$ orbital, the $\gamma$-state to $d_{xz}$, and the $\delta$-state, which is closest to $E_F$, to $d_{yz}$ with the y-axis pointing along the [1$\bar{1}$0] direction (Fig. 1c). Note, that the $\alpha_-$ peak has a strongly increased intensity with respect to its partner $\alpha_+$. This particle-hole asymmetry in the spectral weight of the YSR state is a well-known effect and related to additional spin-independent scattering of the substrate electrons off the corresponding Mn electrons[28,29]. We now turn to the construction of such Mn chains and the investigation of their electronic properties.

**Confined Bogoliubov-de-Gennes quasiparticles in atom-by-atom constructed Mn chains**
The Nb(110) surface allows for STM tip-induced atom manipulation[45] (see Methods). We, thereby, were able to construct atomically well-defined linear chains without any defects (Fig. 1b), which is crucial for enabling the free propagation of quasiparticle waves. The chains are built along the [001] direction consisting of a defined number $N$ of Mn atoms on nearest neighboring four-fold coordinated hollow adsorption sites with a distance of $a$ = 329.4 pm, as sketched in Fig. 1c. All magnetic moments in such Mn$_N$ chains align ferromagnetically, as determined by spin-polarized STM measurements[45]. In order to observe the formation of bands emerging from the YSR states of the Mn atoms, we recorded d$I$/d$V$ spectra along the longitudinal axis through the center of each constructed Mn$_N$ chain (see dashed line in the top panel of Fig. 2a), which are called d$I$/d$V$ line-profiles in the following. The deconvoluted d$I$/d$V$ line-profiles are shown in Fig. 2 for three selected chains of different length (see the data of all chains from $N$ = 1 to $N$ = 36 in Supplementary Note 5). Energetically sharp states with a defined number of LDOS maxima $n_\alpha$ along the chains are formed inside the gap $\Delta_s$. The number of maxima starts from $n_\alpha$ = 1 for the state at an energy of $E_1 \approx +0.5$ meV and then increases in integer steps towards negative energies approaching the coherence peak, where the energy $E_{n_\alpha}$ of these states depends on the length $N$ of the chain. We interpret these states as signatures of QPI of the Bogoliubov-de-Gennes (BdG) quasiparticles confined in the superconductor-magnet-hybrid which we explain further below. As shown in Supplementary Note 5, by plotting the d$I$/d$V$ signal averaged along the entire chain against their length $N$, we can observe a continuous shift of $E_1$ from the energy of the $\alpha_-$ YSR state for $N$ = 1 (i.e. the single atom in Fig. 1d) to $E_1$ = +0.5 meV for $N$ > 10. This provides strong evidence that the corresponding band, which hosts these BdG quasiparticles, is a Shiba band formed by hybridizing $\alpha$-YSR states, and we will consequently name it $\alpha$-YSR band in the following. Most strikingly, there is a region of width $2\Delta_{\text{ind}} \approx 360$ μeV around $E_F$ where the d$I$/d$V$ intensity is strongly reduced, resembling an induced $p$-wave gap in the $\alpha$-YSR band which will be substantiated below.

In addition to the d$I$/d$V$ line-profiles taken along the longitudinal axis of each chain, we mapped the whole d$I$/d$V$ signal in a region covering the chain and its vicinity, which we call d$I$/d$V$ maps in the following. The d$I$/d$V$ maps taken at selected energies of one of the longest chains, Mn$_{34}$, are shown in Fig. 3. The previously discussed states of the $\alpha$-YSR band marked by their numbers of maxima $n_\alpha$ are localized predominantly inside the area on top of the chain which is enclosed by the white dashed lines. We observe additional confined BdG states with a different number of maxima, called $n_\delta$ in the following, which are mainly located outside this area on both sides along the chain and thus have a very weak intensity on the chain's centers. Their maxima in intensity appear slightly offset in the [1$\bar{1}$0] direction from the longitudinal chain axis at a distance and direction that matches the extent and orientation of the lobes of the $\delta$-YSR state observed for the single Mn atoms (see Fig. 1e). It is, thus, very plausible that these states derive from an additional Shiba band formed by the hybridization of the $\delta$-YSR states which we name $\delta$-YSR band in the following.

**Multi-orbital band dispersion extracted from the quasiparticle interference patterns**
In order to extract the dispersion of the $\alpha$- and $\delta$-YSR bands, we analyze the numbers of maxima $n_\alpha$ and $n_\delta$ of the confined BdG quasiparticles in dependence of the energy- and chain length. We recall that the Shiba bands are induced by the local magnetic Zeeman field from the chain and the associated BdG quasiparticles will experience strong confinement in an effective 1D potential well of length $L = Na$ formed by the magnetic chain since they cannot exist in the bulk superconductor. Therefore, the BdG quasiparticles with initial momenta $\mathbf{k}_i$ are scattered at the chain's ends to momenta $\mathbf{k}_f$ by a scattering vector of length $|\mathbf{q}| = |\mathbf{k}_f - \mathbf{k}_i|$. This scattering happens with high probability in the same band. The resulting interference becomes visible as a standing wave in the LDOS($E$), which is approximately proportional to the d$I$/d$V$ signal measured at $V = E/$e. Constructive interference is reached whenever $q = \pm 2n\pi/L = \pm 2n\pi/Na$, with $n \in \mathbb{N}$. By this selection rule, only a discrete set of quasiparticle states of the Shiba bands have non-zero LDOS. The LDOS($E$) then equals $\sum_n |\phi_n|^2 \delta(E - E_n(q))$ with the confined states $\phi_n = \sin(n\pi x/Na)$ (here $x = 0$ corresponds to one of the ends of the chain). We, therefore, fit the d$I$/d$V$ line-profiles of Fig. 2 with a linear combination $dI/dV(E = eV, x) = \sum_{n=0}^{18} c_n(E)|\phi_n(x)|^2$. Unlike the delta-functions, the coefficients $c_n(E)$ take into account the strongly different intensities of the confined BdG quasiparticle states. We restrict the analysis to $n \leq 18$ (see Ref.[46]). They exhibit pronounced peaks at the energetic positions of the confined BdG quasiparticle state with scattering vector $q = \pm 2n\pi/Na$ in the chain and, therefore, relate $E$ and $q$ of the quasiparticles. This method enables us to approximate the dispersion of the scattering vectors $E(q)$ in an infinite chain already for relatively short chain lengths (a similar result is obtained from computing the discrete Fast Fourier Transform of the d$I$/d$V$ line-profiles, see Supplementary Note 4). In Fig.4a, the peaks of $c_n$ for all chains with $14 \leq N \leq 36$ are plotted against the scattering vectors and energy while the colors correspond to the respective peak intensities.

We can identify a dominant nearly parabolic band, which is assigned to the $\alpha$-YSR band, with a band maximum at $E_0 \approx +0.5$ meV ($q$ = 0) and negative effective mass. Note, that we can also resolve the particle-hole partner of this band with a minimum at $-E_0 \approx -0.5$ meV ($q$ = 0) and a positive effective mass, which is compulsory for a BdG quasiparticle band of a superconductor with non-vanishing particle weight, but which was barely visible in the original data representation of Fig. 2. In the region around $E_F$, the $\alpha$-YSR band is gapped by $2\Delta_{\text{ind}} \approx 360$ μeV. Importantly, the intensity of the $\alpha$-YSR band has a comparatively strong particle-hole asymmetry similar to the $\alpha$ YSR state of the single Mn atom (Fig. 1d). Zooming into the gap region close to $q$ = 0, we can also resolve the $\delta$-YSR band which has a Dirac-like dispersion with almost vanishing difference of particle and hole weights. The $\delta$-YSR band crosses $E_F$ at $q_{F,\delta}/2 = \pm 0.1 \pi/a$ without opening a gap within our experimental energy resolution. This band, which is mainly localized alongside the chain (see Fig. 3 and Supplementary Note 4), appears with a weak intensity in the analysis of the d$I$/d$V$ line-profiles which are taken on top of the chains. Still, there is some faint intensity of the corresponding BdG quasiparticles being barely visible in the original data of Fig. 2 within the induced gap. In the following, we will provide evidence of the *p*-wave nature of the gap 2$\Delta_{ind}$ in the $\alpha$-YSR band and possible conclusions about the topological character of this band.

**Tight-binding model for Shiba bands**
We model the Shiba bands by a low-energy model based on Ref.[10] for YSR impurities in a superconducting host with effective Rashba SOC (see Methods). We derive Shiba band structures which closely resemble the experimental $\alpha$- and $\delta$-YSR bands (Fig. 4b,c and Supplementary Figures 6 and 7) using parameters which are consistent (Supplementary Note 6) with the energies and particle-hole asymmetries of the respective single atom YSR states the bands emerge from (Fig. 1d). In particular, Figs. 4b and 4c (left panel) show the according fit to the $\alpha$-YSR band. Originating from the non-degenerate single impurity YSR states (Fig. 1), the $\alpha$-YSR band is non-degenerate itself. A symmetry analysis of the system corroborates this result. We, therefore, conclude that the gap opened in the $\alpha$-YSR band is due to a *p*-wave contribution to the pairing mechanism. From the model calculations, we find that the topological index[1] of the $\alpha$-YSR band is nontrivial. Generally, the topological index can be derived from the experimental data assuming a weak superconducting pairing ($\Delta_{ind}$ << Shiba band width) and if the band features a strong asymmetry in its respective particle-hole weight (Supplementary Note 7), which is both fulfilled for the $\alpha$-YSR band. We extrapolate to the fictitious case of a gapless band structure by setting the effective SOC to zero (dashed line in Fig. 4b) and count the number of Fermi level crossings $c$ = 1 between $k$ = 0 and $k$ = π. This directly relates to Kitaev's[1] $\mathbb{Z}_2$-invariant $M$ = (-1)$^c$ = *-1* for this band. We thus find that the $\alpha$-YSR band carries a non-trivial topological index. In contrast, band structures with trivial gaps (Fig. 4c, center & right panels), which can be simulated by changing the coupling of the individual impurity to the superconductor (parametrized by $A$, see Methods), would have a strongly different appearance in conflict with the experiment.

We suppose that the *p*-wave pairing in the $\alpha$-YSR band is induced by a Rashba-like SOC caused by the breaking of inversion symmetry at the Nb(110) surface[47]. Inversion symmetry breaking naturally has a stronger effect on the $d_{z^2}$-orbitals pointing along the surface normal as compared to the in-plane oriented $d_{yz}$-orbitals[48], which could explain the absence of a measurable gap in the $\delta$-YSR band. In addition, the superconducting order parameter of a Shiba chain in the presence of SOC becomes *k*-dependent and thus can prevent the opening of a measurable gap at $k$ values where $\Delta_k$ vanishes (Supplementary Note 6 and Supplementary Figure 7). We argue, however, that it is unlikely that the *p*-wave pairing in the $\delta$-YSR band is exactly zero but rather just too small to be experimentally resolved.

**Discussion**
We finally discuss the possible emergence of topological edge states from the observed Shiba bands. MBS are, in principle, expected to form at the chain's ends, if there is an odd number of topological bands, all bands are gapped and if the chain is long compared to the localization length of the MBSs $\xi_M$[11,49]. The question arises whether the zero-energy mode observed in Fig. 3 could be the MBS of the $\alpha$-YSR band. Zero-energy states with a similar spatial distribution have been attributed to topological MBSs in previous work[18]. However, in the system investigated here these states clearly do not belong to the $\alpha$-YSR band since MBS arising from this band would have a modulation of $q_{F,\alpha}/2 \approx 0.2$ π/$a$, i.e. equal to the scattering vector at the gap minimum, imprinted on their LDOS[49,50] (see also Supplementary Figure 6b). The zero-energy state in Fig. 3 is modulated with only about half of this wavevector and we can therefore exclude, that it is the zero energy MBS stemming from the $\alpha$-YSR band. Instead, since the $\delta$-YSR band remains gapless up to the experimental resolution, some confined BdG quasiparticles of this band are usually located inside the gap $\Delta_{ind}$ of the $\alpha$-YSR band. For certain lengths of the chain, they can be located very close to $E_F$, as, e.g., the state with $n_\delta$ = 3 of the Mn$_{34}$ chain marked in the d$I$/d$V$ maps in Fig. 3.

Using our knowledge about the $\alpha$-YSR band dispersion, we are able to estimate the expected MBS localization length via $\xi_M = \hbar \tilde{v}_F/\Delta_{ind}$ from the experimental value of the induced gap $\Delta_{ind} \approx 180 \mu eV$ and the Fermi velocity $\tilde{v}_F$, which is evaluated from the slope of the $\alpha$-YSR band close to $E_F$ via $\tilde{v}_F = 2\hbar^{-1} \partial E/\partial q \, (q_F) \approx 1 \cdot 10^3$ m/s, yielding $\xi_M \approx 3.8$ nm (see Supplementary Note 4 and Supplementary Figure 4). Therefore, in our longest chains ($N = 36$) of length $L = Na \approx 12$ nm the expected MBS of the $\alpha$-YSR band should have only little overlap. However, their apparent absence in our data would be explained by hybridization with states of the $\delta$-YSR band or with states of potentially experimentally invisible Shiba bands formed by the hybridizing YSR states of the other orbitals. We emphasize that the presence of additional unresolved bands may change the total topological phase of the chain, which is the product of the individual topological indices of all relevant bands. We cannot experimentally exclude that additional topologically non-trivial bands originating from $d_{xy}$-, $d_{xz}$-, or $d_{x^2-y^2}$-orbitals exist, as tunneling into these in-plane orbitals is strongly suppressed (Fig. 1d) which explains the invisibility of these bands in Fig. 4. Hybridizations between such bands generally destroy the topological protection of a MBS apart from very special conditions[51].

In conclusion, our experimental results represent the first observation of multi-orbital Shiba band dispersions in a magnet-superconductor hybrid system. We find at least two bands of sufficient band width to possibly undergo a topological band inversion and we can address these bands individually by properly choosing the tip position relative to the magnetic chain. One of the two bands is topologically gapped around the Fermi level, indicating that the Rashba SOC due to breaking of inversion symmetry at the (110) surface of Nb is sufficient to induce p-wave correlations in ferromagnetic bands. The absence of zero energy modes in this band implies additional constraints for realizing MBSs in the multi-orbital magnetic chain on superconductor platform. Resolving the entire band structure of similar hybrid systems can enable the determination of their overall topological phase by bulk properties only, since the origin of the topological boundary states, such as MBS, is encoded in the system's bulk bands. Therefore, a proper understanding and possibly even controlled manipulation of the sub-gap band structure may enable the design of topological superconductors in future experiments.

## Methods

### Experimental procedures

All experiments have been performed in a home-built low-temperature STS facility at a base temperature of $T = 0.32$ K[43]. A clean Nb(110) surface was obtained by flashing the single crystal to $T > 2700$ K[41]. Single Mn atoms were deposited successively while keeping the substrate cooled below $T = 7$ K. This results in a statistical distribution of the adatoms on the surface, as shown in Fig. 1a.

STM images were obtained maintaining a constant-current $I$ while applying a bias voltage $V$ across the tunneling junction. For the measurement of d$I$/d$V$($V$) spectra, the tip was stabilized at bias voltage $V_{stab}$ and current $I_{stab}$. Subsequently, the feedback loop was opened and the bias voltage was swept from -4 mV to +4 mV. The differential tunneling conductance d$I$/d$V$ was measured using a standard lock-in technique with a small modulation voltage $V_{mod}$ (rms) of modulation frequency $f$ = 4.142 kHz added to the bias voltage. The d$I$/d$V$ line-profiles and maps were acquired recording multiple d$I$/d$V$ spectra along a line or grid, respectively. One-dimensional chains were assembled using lateral atom manipulation techniques at low tunneling resistances of $R \approx 30 - 60$ kΩ.

### Tight-binding model for Shiba bands

Pientka *et al.* developed a model[10] for a chain of YSR impurities embedded in a superconducting substrate, which describes weakly interacting impurities at energies close to the Fermi level. We extend the model to include single-particle scattering by a YSR impurity. This is an essential step to describe the experimentally observed particle-hole asymmetry.

The BdG Hamiltonian of a chain consisting of $n$ YSR impurities that are spin-polarized out-of-plane is

$$H = \epsilon_p \tau^z + \Delta_s \tau^x + \sum_{j=1}^{n} (V\tau^z - J\sigma^z) \delta(\mathbf{r} - j\, a\, (1,0,0)^T) \quad (1)$$

where $a$ is the distance between the impurities, $\Delta_s$ denotes the superconducting s-wave pairing, $J$ is the exchange coupling between the magnetic impurity and the superconductor, $V$ denotes the non-magnetic scattering at the impurity and $\tau^i = \mathbf{1}_{2x2} \otimes s^i$ and $\sigma^i = s^i \otimes \mathbf{1}_{2x2}$ (with Pauli matrices $s^i$) act on the Nambu space in the basis $(c_\uparrow^\dagger(\mathbf{r}), c_\downarrow^\dagger(\mathbf{r}), c_\downarrow(\mathbf{r}), -c_\uparrow(\mathbf{r}))$, and $\otimes$ is the Kronecker product. The dispersion of the superconductor $\epsilon_p = (mp^2/2 - \mu) + i\lambda(p_x\sigma^y - p_y\sigma^x)$ includes Rashba SOC[47] quantified by $\lambda$, which is experimentally motivated by the breaking of the z-inversion symmetry at the surface of the substrate, where the chain is located. The results of Ref.[10] are recovered for $V = 0$ and by noting that Rashba SOC is equivalent to a magnetic helix[52] with an angle $\phi = 2\lambda ma/\hbar$ between adjacent spins, described by the pitch $\pi/k_h$ with $k_h = \phi/(2a) = \lambda m/\hbar$. To obtain the effective tight-binding model for the Shiba bands, we have followed the lines of Ref.[10], which we refer to for details.

For the low-energy model of the Shiba band, we obtain the BdG Hamiltonian with on-site potential $h_{i,i}$, hopping $h_{i,j}$ ($i \neq j$), and effective *p*-wave superconductivity $\Delta_{i,j}$ ($i \neq j$)

$$h_{i,i} = \Delta_s \frac{\left(A - \sqrt{(A^2 - B^2)^2 + B^2}\right)}{(A-B)(A+B)}, \tag{2}$$

$$h_{i,j} = -\frac{\Delta_s \, e^{\frac{-a|i-j|}{\xi}} \cos[k_h \, a \, (i-j)] \left( m_{1,1}(A,B) \cos[k_F \, a \, |i-j|] + m_{1,2}(A,B) \sin[k_F \, a \, |i-j|] \right)}{k_F \, a \, |i-j|}, \tag{3}$$

$$\Delta_{i,j} = -\frac{\Delta_s \, e^{\frac{-a|i-j|}{\xi}} \sin[k_h \, a \, (i-j)] \left( m_{2,1}(A,B) \cos[k_F \, a \, |i-j|] + m_{2,2}(A,B) \sin[k_F \, a \, |i-j|] \right)}{k_F \, a \, |i-j|}. \tag{4}$$

Here $A = \pi v_0 J$, and $B = \pi v_0 V$, where $v_0$ is the normal-phase density of states[10], $k_F$ is the Fermi wavevector and $\xi$ is the effective coherence length in the Shiba chain. The coefficients $m_{1,1}, m_{1,2}, m_{2,1}$, and $m_{2,2}$ for general $A$ and $B$ which have been used for the fits are given in Supplementary Note 6. When $A^2 = 1 + B^2 + \epsilon$, with a small $\epsilon$, we reach at

$$m_{1,1}(A,B) = -m_{2,2}(A,B) = \frac{B^2}{\sqrt{1+B^2}}. \tag{5}$$

and

$$m_{1,2}(A,B) = m_{2,1}(A,B) = \frac{1 + 2B^2(1-\epsilon)}{\sqrt{1+B^2}}. \tag{6}$$

up to first order in $\epsilon$. The effective Hamiltonian in Eqs. (2-4) acts on the basis states of the YSR states with a particle-weight[53]

$$P(A,B) = \frac{1 + (A+B)^2}{2(1+A^2+B^2)}. \tag{7}$$

To obtain the particle-hole asymmetry of a state in terms of the physically original particles, which can be extracted from the particle-hole asymmetry in the spectral weight in Fig. 1d, $P(A,B)$ in Eq. 7 is multiplied with the particle-weight of a state and $(1 - P(A,B))$ is multiplied with the hole-weight.

For an infinite chain, we obtain the band structure by transforming the Hamiltonian to Fourier space[10]. This yields an analytic formula, which we employed for determining the effective coupling parameters by numerical fits to the experimental data (see Fig. 4b,c and Supplementary Note 6).

**Data availability**
The authors declare that the data supporting the findings of this study are available within the paper and its supplementary information files.

**Code availability**
The analysis codes that support the findings of the study are available from the corresponding authors on reasonable request.

**Acknowledgements**
L.S., P.B., T.P., J.W., and R.W. gratefully acknowledge funding by the Cluster of Excellence 'Advanced Imaging of Matter' (EXC 2056 - project ID 390715994) of the Deutsche Forschungsgemeinschaft (DFG). L.S., P.B., J.W., and R.W. acknowledge support by the SFB 925 'Light induced dynamics and control of correlated quantum systems' of the Deutsche Forschungsgemeinschaft (DFG). R.W. acknowledges funding of the European Union via the ERC Advanced Grant ADMIRE. S.R. acknowledges support from the Australian Research Council (DP200101118). We thank Dirk K. Morr, Harald Jeschke and Levente Rózsa for helpful discussions. T.P. thanks Falko Pientka for clarifying remarks.

**Author contributions**
L.S., P.B., R.W. and J.W. conceived the experiments. L.S. and P.B. performed the measurements and analyzed the experimental data together with J.W.. T.P. derived the effective low energy Shiba model. L.S. performed the numerical simulations using the effective low energy Shiba model and the fitting to the experimental data. D.C., E.M. and S.R. performed numerical simulations which were essential for the understanding of the system. L.S. prepared the figures. L.S. and J.W. wrote the manuscript. All authors contributed to the discussions and to correcting the manuscript.

**Competing interests**
The authors declare no competing interests.

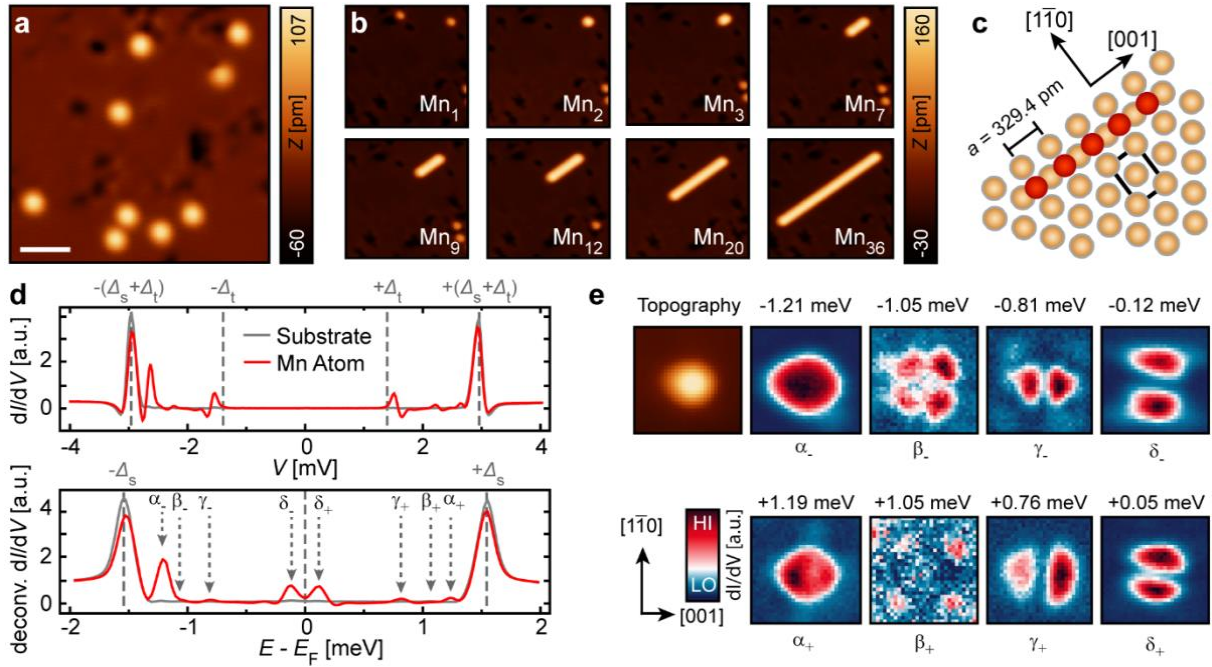

**Figure 1 | Multi-orbital YSR states of Mn atoms and construction of Mn chains. a**, STM image of single Mn atoms adsorbed on a clean Nb(110) surface. The white bar corresponds to 2 nm. **b**, Series of STM images obtained during the construction process of $Mn_N$ chains (*N* as indicated) using atom manipulation. **c**, Sketch of the uppermost layer of Nb atoms (brown) and the positions of Mn atoms (red) arranged in a linear chain along the [001] direction. The lattice constant $a$ of Nb is indicated. **d**, d$I$/d$V$ spectra taken using a superconducting tip averaged over the center of a single Mn atom and over a clean area of the substrate (upper panel). The lower panel shows the same data after numerical deconvolution. The energetical positions of the sample ($\Delta_s$) and tip ($\Delta_t$) gaps are indicated by dashed vertical lines and of the pairs of multi-orbital YSR peaks by Greek letters. **e**, d$I$/d$V$ maps around a single Mn atom shown in the STM image (topography) evaluated at selected bias voltages according to the YSR-multiplet energies. The panels are 2x2nm² each. Parameters: $V$ = 50 mV, $I$ = 50 pA for **a**, $V$ = -6 mV, $I$ = 200 pA for **b** and $V_{stab}$ = -6 mV, $I_{stab}$ = 1 nA, $V_{mod}$ = 20 μV for **d** and **e**.

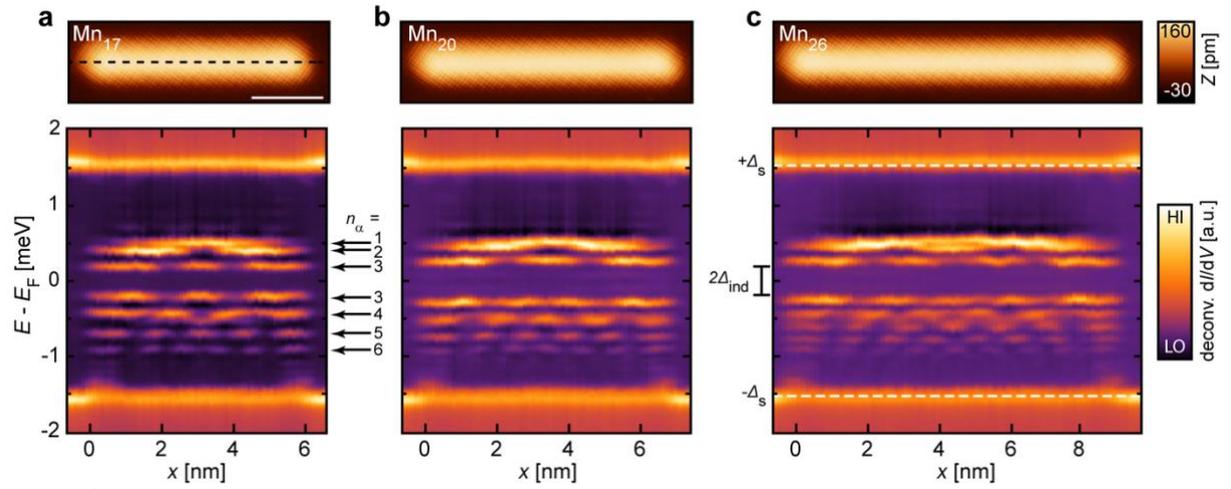

**Figure 2 | BdG quasiparticle interference patterns at the center of the Mn chains. a, b, c**, The upper panels show STM images of Mn$_N$ chains containing $N$ Mn atoms as indicated. The white bar corresponds to 2 nm. The lower panels show deconvoluted d$I$/d$V$ spectra ("d$I$/d$V$-line-profiles") taken along the longitudinal axis through the center of each chain as indicated by the dashed line in **a**, being aligned with the upper panels. The energetical positions of the sample gap ($\Delta_s$) and of the induced gap ($\Delta_{ind}$) are indicated by dashed horizontal lines and the bar in **c**, respectively. The number of maxima $n_\alpha$ of BdG confined quasiparticles is indicated in the bottom panel of **a**. Parameters: $V_{stab}$ = -6 mV, $I_{stab}$ = 1 nA, $V_{mod}$ = 20 µV.

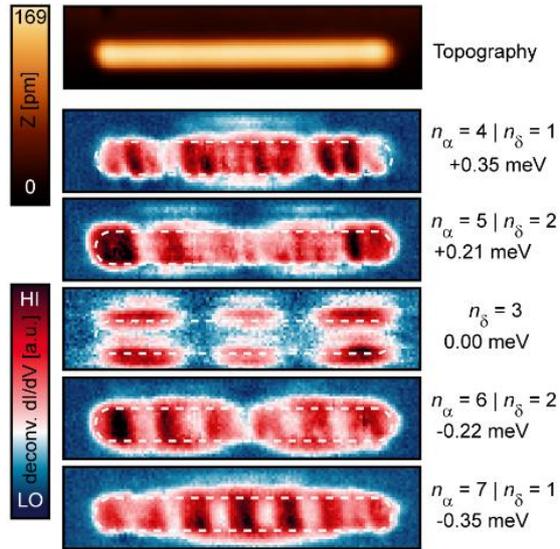

**Figure 3 | 2D maps of BdG quasiparticle interference patterns on the Mn chains.** STM image (top panel) and d$I$/d$V$ maps (bottom panels) taken on a Mn$_{34}$ chain at selected energies as indicated. The location of the chain is marked in the d$I$/d$V$ maps by the white dashed lines. The numbers of maxima of the $\alpha$- and $\delta$-YSR bands $n_{\alpha/\delta}$ are indicated. Parameters: $V_{stab}$ = -6 mV, $I_{stab}$ = 1 nA, $V_{mod}$ = 20 µV.

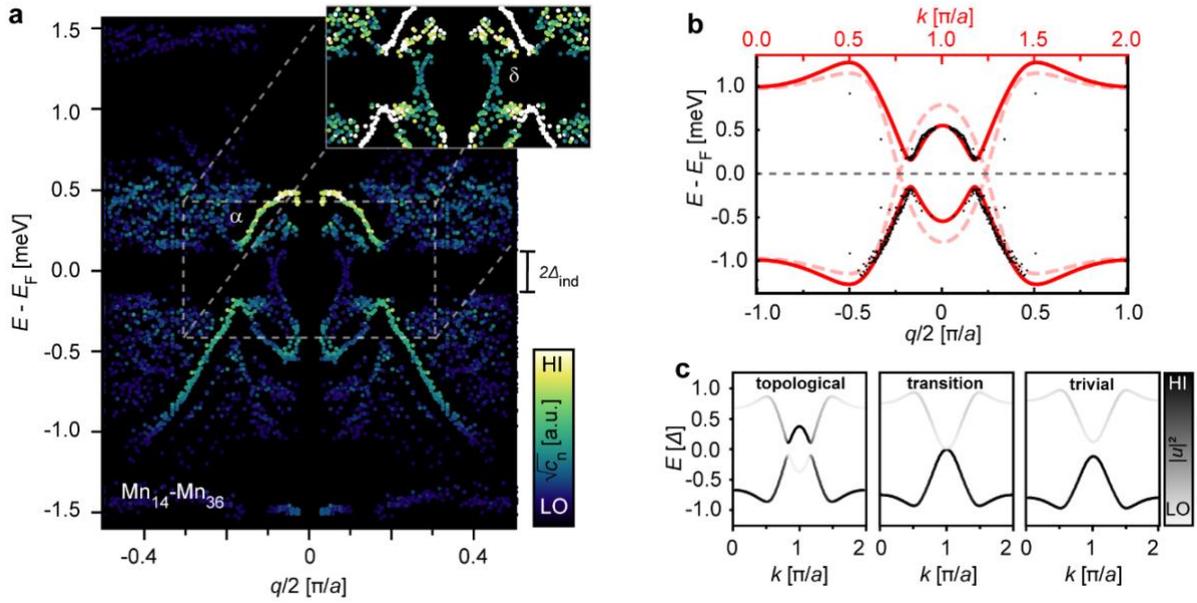

**Figure 4 | Dispersion of Multi-orbital BdG quasiparticle scattering vectors. a**, Strongest peaks in the coefficients $c_n$ plotted against their energies and scattering vectors $q/2 = n\pi/(Na)$ extracted from the d$I$/d$V$ line-profiles of Fig. 2 of all chains with $14 < N < 36$. The color indicates the respective intensities of the peaks. The induced gap $\Delta_{ind}$ is indicated on the right side of the panel. The inset shows the central region with adjusted contrast. The α- and δ-YSR bands are indicated by the Greek letters. **b**, Fit of the band dispersion from the Shiba chain model (solid line) to the manually evaluated experimental data of the α-YSR band (dots, Supplementary Note 4). Parameters of the model: $\Delta_s$ = 1.5 meV, $A$ = 3.1, $B$ = 2.35, $\xi$ = 0.77 nm, $k_F$ = 0.69 π/a, $k_h$ = 0.14 π/a. For the dashed line the $p$-wave gap was artificially set to zero ($k_h$ = 0). **c**, Experimentally detectable particle contributions of the band dispersions from the Shiba chain model for the topological phase (left) using the same parameters which fit to the experimental data in **a** and **b**, at the topological phase transition using $A$ = 3.6 (center) and in a trivial phase using $A$ = 3.9 (right).